\documentclass[11pt,twoside]{article}
\usepackage{asp2010}

\resetcounters

\begin{document}

\title{Spiral Structure Dynamics in Pure Stellar Disk Models}
\author{Diego Valencia-Enr\'iquez$^1$ and Iv\^anio Puerari$^{1,2}$}
\affil{$^1$Instituto Nacional de Astrof\'isica, \'Optica y Electr\'onica,\\
Santa Mar\'ia Tonantzintla, Puebla, M\'exico.}
\affil{$^2$IP\&D, UniVap, S\~ao Jos\'e dos Campos, Brazil.}

\begin{abstract}
In order to understand the physical mechanism underlying non-steady stellar spiral arms in disk galaxies we performed a series of N-body 
simulations with 1.2 and 8 million particles. The initial conditions were chosen to follow Kuijken-Dubinski models. In this work we present 
the results of a sub-sample of our simulations in which we experiment with different disk central radial velocity dispersion ($\sigma_{R,0}$)
and the disk scale height ($z_d$). 

We analyzed the growth of spiral structures using 1D and 2D Fourier Transform (FT1D and FT2D respectively). The FT1D was used to obtain 
the angular velocities of non-axisymmetric structures which grow in the stellar disks. In all of our simulations the measured angular velocity 
of spiral patterns are well confined by the resonances given by the curves $\Omega \pm \kappa/m$. The FT2D gives the amplitude of a particular 
spiral structure represented by two Fourier frequencies: $m$, number of arms; and $p$, related to the pitch angle as $atan(-m/p)$. We present, 
for the first time, plots of the Fourier amplitude $|A(p,m)|$ as a function of time which clearly 
demonstrates the swing amplification mechanism in the simulated stellar disks. In our simulations, the spiral waves appear as leading spiral
structures evolving towards open trailing patterns and fade out as tightly wound spirals.
%This is the Astronomical Society of the Pacific (ASP) 2010 author template file.  This sample author template includes the ASP author checklist.
\end{abstract}

\section{Introduction, simulations and results for 1D and 2D Fourier Transform}
The physical origin and evolution of spiral arms in disk galaxies is a long-standing problem in Galactic astronomy. 
The most widely accepted hypothesis is the Spiral Density Wave theory \citep{1964ApJ...140..646L} %(Lin & Shu, 1964) 
 which explains that the spiral arms are long lived quasi-stationary density waves. However, \cite{1981seng.proc..111T} %Toomre (1981) 
proposed short lived arms. In his mechanism, a short leading spiral perturbation shears at corotation into a short 
trailing spiral due to differential rotation. The wave is amplified by self-gravity due to assembling of 
stars in the perturbation, and moreover at the expense of some disk rotational energy.

The initial conditions of our simulations were chosen to follow \cite{kuijken1995} %Kuijken & Dubinsky (1995)
models. We reproduce one of their models with disk central radial velocity dispersion $\sigma_{R,0}=0.47$ and disk scale height 
$z_d=0.10$, and created 16 models with 1.2 million particles each combining $\sigma_{R,0} = [0.27,\ 0.37,\ 0.47,\ 0.57]$ and $z_d = [0.05,\ 0.10,\ 0.15,\ 0.20]$. Furthermore, we created 6 models with 8 million particles each combining 
$\sigma_{R,0} = [0.27,\ 0.37]$ and $z_d = [0.05,\ 0.10,\ 0.15]$. All the models were evolved in a time range from 0 to 5 Gyr.

\begin{figure}[!ht]
\centering
 \includegraphics[scale=0.58]{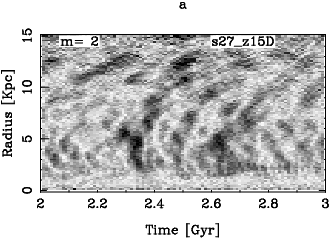}
 \includegraphics[scale=0.58]{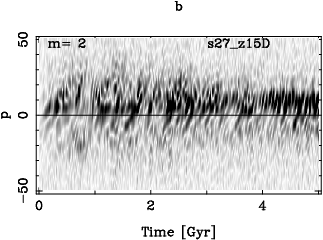}
 \includegraphics[scale=0.58]{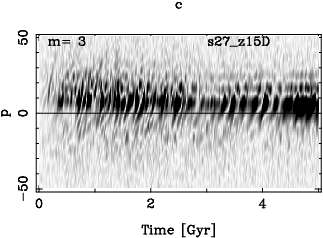}
 \includegraphics[scale=0.58]{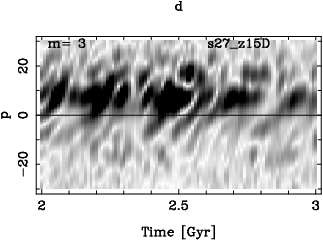}
 \caption{(a) FT1D zoom for one of our models for $m=2$. (b) FT2D for $m=2$ for the same model of figure (a). (c) As in (b), but for $m=3$. (d) FT2D zoom for $m=2$.}
 \label{valenciad_figs}
\end{figure}

In order to understand the growth of spiral patterns in the disk, we have performed the FT1D in polar coordinates $(R,\ \phi)$. 
Figure (\ref{valenciad_figs}a) shows a zoom of the FT1D amplitude as function of radius and time for $m=2$ for a particular model. 
The black regions are over-densities where the perturbation appears, these appear close to a radius of 6 kpc in the disk 
and grow towards inner and outer parts of the disk. \textit{The perturbation appears where the criterion of stability $Q$ 
is minimum and these patterns are more intense for colder and thinner models}. The angular velocity of spiral patterns that we have 
measured from our simulations is well confined between the main resonances given by the curves $\Omega \pm \kappa/m$. 
This behavior is observed in all of our simulations.

We used the FT2D method, as described in \cite{1992A&AS...93..469P}, %Puerari & Dottori (1992), 
to obtain the theoretical mechanism that best resembles our simulated particle disks.  
Figures (\ref{valenciad_figs}b) and (\ref{valenciad_figs}c) show the FT2D amplitude for $m=2$ and $m=3$ as a function of 
$p$ and time for one of the models. In these figures each black region is a time evolution of a spiral pattern. 
Trailing structures appear at positive $p$ and leading structures at negative $p$.  \textit{It is clear that spiral 
pattern evolves from leading to trailing evidencing of the swing amplification mechanism}. All of our models 
follow this behavior. Figure (\ref{valenciad_figs}c) shows that the presented model developed multi-arm spirals. These patterns 
are more intense for hotter and thicker models, i.e., while thin/cold models develop strong $m=2$ patterns, thick/hot ones tend
to develop spiral structures with higher azimuthal frequencies (higher m's). In figure (\ref{valenciad_figs}d) we can observe that the life time of 
the spiral patterns is about 300 million years, and that they appear every 100 million years. This can be explained as \textit{being the 
spiral patterns a superposition of three or more waves with different angular velocities which made them 
grow and decay later on} \citep{2011MNRAS.410.1637S}.%(Sellwood, 2011). 

When comparing the models with 8 million particles with those of 1.2 we found no significant differences. The main effect coming for enlarging the number 
of particles is that the models with larger mass resolution develop stronger spiral patterns at later times.

\bibliography{valenciad.bib}

\end{document}